# Large Spin Pumping from Epitaxial $Y_3Fe_5O_{12}$ Thin Films to Pt and W Layers

H. L. Wang[†], C. H. Du[†], Y. Pu, R. Adur, P. C. Hammel[*], and F. Y. Yang[*]

Department of Physics, The Ohio State University, Columbus, OH, 43210, USA

[†]These authors made equal contributions to this work

[*]Emails: hammel@physics.osu.edu; fyyang@physics.osu.edu

Abstract

Epitaxial $Y_3Fe_5O_{12}$ thin films have been deposited by off-axis sputtering, which exhibit excellent crystalline quality, enabling observation of large spin pumping signals in Pt/$Y_3Fe_5O_{12}$ and W/$Y_3Fe_5O_{12}$ bilayers driven by cavity ferromagnetic resonance. The inverse spin Hall voltages reach 2.10 mV and -5.26 mV in 5-mm long Pt/$Y_3Fe_5O_{12}$ and W/$Y_3Fe_5O_{12}$ bilayers, respectively, excited by a radio-frequency magnetic field of 0.3 Oe. From the ferromagnetic resonance linewidth broadening, the interfacial spin mixing conductance of $4.56 \times 10^{14}$ $\Omega^{-1}m^{-2}$ and $2.30 \times 10^{14}$ $\Omega^{-1}m^{-2}$ are obtained for Pt/$Y_3Fe_5O_{12}$ and W/$Y_3Fe_5O_{12}$ bilayers, respectively.

Ferromagnetic resonance (FMR) driven spin pumping of pure spin currents has generated intense interest for its potential application in next-generation spintronics [1-17]. Due to the exceptionally low magnetic damping, $Y_3Fe_5O_{12}$ (YIG) has been regarded as one of the best ferromagnets (FM) for microwave applications and FMR spin pumping [1-9]. The inverse spin Hall effect (ISHE) is an effective tool for studying spin pumping from FMs into nonmagnetic materials (NM) [1-4, 12, 14, 15]. In addition to Pt which is widely used as a NM due to its large ISHE, $\beta$-phase W and Ta are expected to generate large ISHE voltages (though of the opposite sign), making them attractive in this role as well. To date, no clear ISHE detection of FMR spin pumping in W/FM structures has been reported. Generating a high spin current density with a modest radio-frequency (rf) field, $h_{rf}$, requires a FM with low damping and YIG is highly attractive for this purpose [18]. In this letter, we report observation of ISHE voltages, $V_{ISHE}$, of 2.10 mV (0.420 mV/mm) and 5.26 mV (1.05 mV/mm) for Pt/YIG and W/YIG bilayers, respectively, excited by a rf field of 0.3 Oe in a FMR cavity.

There are two common methods in generating magnetic resonance in FMs for spin pumping, cavity FMR and microstrip waveguides [3, 7, 8, 12, 19]. FMR cavities produce modest-strength, uniform rf fields over a relatively large space (cm-scale); while microstrip waveguides produce rf fields typically in micron to sub-mm scale, and when made very close to the FMs, can generate fairly large $h_{rf}$ [12, 19]. Since the magnitude of rf field determines the excitation strength for spin pumping and only a few reports on microstrip spin pumping presented values of $h_{rf}$ [12, 19], in this letter, we mainly compare our results with previous reports of spin pumping using cavity FMR.

Most YIG epitaxial films and single crystals are produced by liquid-phase epitaxy (LPE) with thicknesses from 100 nm to millimeters [20]. Pulsed laser deposition (PLD) has also been used to grow epitaxial YIG thin films [21-23], although no ISHE measurement of spin pumping is reported. Using a new approach of ultrahigh vacuum off-axis sputtering [24-26], we deposit epitaxial YIG thin films on (111)-oriented $Gd_3Ga_5O_{12}$ (GGG) substrates (see supplementary information for details).

The crystalline quality of the YIG films is determined by high-resolution x-ray diffraction (XRD). A representative $\theta$-$2\theta$ scan of a 20-nm YIG film in Fig. 1a indicates a phase-pure epitaxial YIG film. Figure 1b shows $\theta$-$2\theta$ scans near the YIG (444) peak for four films with thicknesses, $t = 10, 20, 50$ and $80$ nm, from which the out-of-plane lattice constant of the YIG films are obtained: $c = 12.426$ Å, $12.393$ Å, $12.383$ Å and $12.373$ Å, respectively. Except for the 10-nm film, all other YIG films have lattice constants very close (within 0.14%) to the bulk value of 12.376 Å, indicating essentially strain-free films. Pronounced Laue oscillations are observed in all films, reflecting smooth surfaces and sharp YIG/GGG interfaces. The XRD rocking curves (insets to Fig. 1b) exhibit a full width at half maximum (FWHM) of 0.027°, 0.0092°, 0.0072°, and 0.0053° for the 10, 20, 50, and 80 nm thick films, respectively, which reach the resolution limit of conventional high-resolution XRD systems, demonstrating excellent crystalline quality. In this letter, we focus on two 20-nm YIG films (YIG-1 and YIG-2) for FMR and spin pumping measurements.

Room-temperature FMR measurements of the YIG films are carried out in a cavity at a microwave frequency $f = 9.65$ GHz and power $P_{rf} = 0.2$ mW. Figure 2 shows an FMR derivative spectrum of a 20-nm YIG film (YIG-1) with an in-plane magnetic field $H$ along the $x$-axis ($\theta_H = 90°$, see top-right inset to Fig. 2 for FMR measurement geometry), which gives a peak-to-peak linewidth ($\Delta H$) of 7.4 Oe (for YIG-2, $\Delta H$ = 11.7 Oe). The angular dependence of the resonance field ($H_{res}$) of the YIG film is shown in the bottom-left inset to Fig. 2b, where $H_{res}$ is defined as the field at which the derivative of the FMR absorption crosses zero. We obtain the effective magnetization, $4\pi M_{eff}$ = 1794 Oe, from a fit to $H_{res}(\theta_H)$ employing quantitative analysis [27, 28], which agrees well with the values reported for single crystal YIG [29].

Our spin pumping measurements are conducted at room temperature on three bilayer samples: Pt(5nm)/YIG-1, Pt(5nm)/YIG-2 and $\beta$-W(5nm)/YIG-2, all made by off-axis sputtering. The samples with approximate dimensions of 1 mm × 5 mm are placed in the center of the FMR cavity with $H$ applied in the $xz$-plane while the ISHE voltage is measured across the 5-mm long Pt or W layer along the $y$-axis, as illustrated in Fig. 3a. The transfer of angular momentum to the Pt or W conduction electrons [30, 31] resulting from FMR excitation of the YIG magnetization ($M$) can be described as a spin current $J_s$ injected along the $z$-axis with its polarization ($\sigma$) parallel to $M$. This spin current is converted by spin-orbit interactions to a charge current $J_c \propto \theta_{SH} J_s \times \sigma$, where $\theta_{SH}$ is the spin-Hall angle of Pt or W [32]. Figure 3b shows the $V_{ISHE}$ vs. $H$ spectra for Pt/YIG-1 and W/YIG-2 at $\theta_H = 90°$ (field in-plane) and $P_{rf}$ = 200 mW, which generates an rf field $h_{rf}$ ~0.3 Oe. At this moderate $h_{rf}$ excitation, $V_{ISHE}$ reaches a large value of 1.74 mV (0.35 mV/mm) in Pt/YIG-1, significantly larger than previously reported spin pumping signals using cavity FMR [1, 6, 9-11, 13-16].

The W/YIG-2 bilayer exhibits an even larger $V_{ISHE}$ of -5.26 mV (-1.05 mV/mm), where the negative sign reflects the opposite spin Hall angles of W and Pt [33].

Figure 3c shows the rf-power dependence of $V_{ISHE}$ for Pt/YIG-1 and W/YIG-2 at $\theta_H = 90°$. The linear relationship between $V_{ISHE}$ and $P_{rf}$ indicates that the observed ISHE voltage is not near saturation and can potentially be further increased by larger $h_{rf}$ (~0.3 Oe in our measurements) since $V_{ISHE} \propto (h_{rf})^2$ [19]. Figure 3d shows a series of $V_{ISHE}$ vs. $H$ spectra for varying $\theta_H$ at $P_{rf} = 200$ mW for the two samples. $V_{ISHE}$ vs. $H$ is antisymmetric about $H = 0$ as expected from FMR spin pumping since the reversal of **H** switches **M** (hence **σ**) and, consequently, changes the sign of **J_c**. When **H** is rotated from in-plane to out-of-plane, $V_{ISHE}$ gradually vanishes. **M** approximately follows **H** at all angles since 2500 Oe $< H_{res} <$ 5000 Oe, all larger than $4\pi M_{eff} = 1794$ Oe of our YIG film. Figure 3e shows the angular dependence of $V_{ISHE}$ for Pt/YIG-1 and W/YIG-2 normalized by the maximum magnitude of $V_{ISHE}$ at $\theta_H = 90°$. The clear sinusoidal shape is characteristic of ISHE since [15]

$$V_{ISHE} \propto \boldsymbol{J_c} \propto \theta_{SH}\boldsymbol{J_s}\times\boldsymbol{\sigma} \propto \theta_{SH}\boldsymbol{J_s}\times\boldsymbol{M} \propto \theta_{SH}\boldsymbol{J_s}\times\boldsymbol{H} \propto \theta_{SH}\sin\theta_H, \quad (1)$$

thus confirming that the observed ISHE voltage arises from FMR spin pumping. The spin pumping signals we observed in insulating YIG cannot be explained by artifacts due to thermoelectric or magnetoelectric effects, such as anisotropic magnetoresistance (AMR) or anomalous Hall effect (AHE) [13, 16, 32, 34, 35].

While a spin current is generated by transfer of angular momentum from YIG to metal, simultaneously, the coupling between YIG and metal exerts an additional damping to the magnetization precession in YIG, resulting in increased linewidths [10, 12], as shown in Fig.

4 for the three samples before ($\Delta H_0$) and after ($\Delta H_1$) the deposition of Pt or W. A clear linewidth broadening is observed for all three samples: $\Delta H_1 - \Delta H_0$ = 19.9, 24.3 and 12.3 Oe for Pt/YIG-1, Pt/YIG-2 and W/YIG-2, which give $V_{ISHE}$ of 1.74, 2.10 and 5.26 mV, respectively. We note that the magnitude of $V_{ISHE}$ appears more correlated to the linewidth change than the original linewidths of the YIG films: Pt/YIG-2 has larger linewidth increase (24.3 Oe) and $V_{ISHE}$ (2.10 mV) than Pt/YIG-1 ($\Delta H_1 - \Delta H_0$ = 19.9 Oe, $V_{ISHE}$ = 1.74 mV) although YIG-2 ($\Delta H_0$ = 11.7 Oe) has a larger linewidth than YIG-1 ($\Delta H_0$ = 7.4 Oe). This can be understood that while narrower FMR linewidth leads to a larger FMR cone angle, the linewidth change determines the interfacial spin mixing conductance which is critically important for spin pumping efficiency [10, 12],

$$G_r = \frac{e^2}{h} \frac{2\sqrt{3}\pi M_s \gamma t_F}{g\mu_B \omega} (\Delta H_1 - \Delta H_0), \qquad (2)$$

where $G_r$, $\gamma$, $g$ and $\mu_B$ are the real part of spin mixing conductance, the gyromagnetic ratio, $g$ factor and Bohr magnetron, respectively. Using Eq. (2), we obtain $G_r$ = 4.56× $10^{14}$ and 2.30× $10^{14}$ $\Omega^{-1}m^{-2}$ for Pt/YIG-2 and W/YIG-2, respectively, which agree with the theoretical calculations [36] and are among the highest of reported experimental values [3, 5, 8, 9].

Previously, spin pumping of Pt/YIG excited by similar cavity FMR gave ISHE voltages in the μV range [1, 9, 11, 16]. The large spin pumping signals and high spin mixing conductance observed in our YIG films may be attributed to two possible reasons. First, the small thickness (20 nm) of our films compared to LPE films (100 nm or larger) may play an important role, as suggested by a recent report [7] that a 200-nm YIG film shows much higher spin pumping efficiency than 1-μm and 3-μm films excited by a microstrip waveguide.

Secondly, the YIG films made by our off-axis sputtering method may be different in crystalline quality and FMR characteristics from those by other techniques. Compared to cavity FMR, microstrip waveguides can potentially provide much stronger rf fields, e.g. 16 Oe in Ref. 19 and 4.5 Oe in Ref. 12, which can significantly increase the magnitude of ISHE voltages ($V_{ISHE} \propto h_{rf}^2$ in the linear regime) [7, 12]. Further investigation of spin pumping in these thin YIG films using microstrip waveguides will access larger dynamic range of spin pumping. In addition, the mV-level ISHE voltages reported here using a moderate $h_{rf}$ will allow miniaturization of spin pumping structures while maintaining signals sufficiently large to explore opportunities such as magnon-based electronics and other next generation technologies [18]. It also provides a material platform for probing the fundamental mechanisms in spin pumping for quantitative characterization of coupling mechanisms and interfacial phenomena.

**Acknowledgements**

This work is supported by the Center for Emergent Materials at the Ohio State University, a NSF Materials Research Science and Engineering Center (DMR-0820414) (HLW, YP, and FYY) and by the Department of Energy through grant DE-FG02-03ER46054 (RA, PCH). Partial support is provided by Lake Shore Cryogenics Inc. (CHD) and the NanoSystems Laboratory at the Ohio State University.

**Figure Captions:**

**Figure 1.** (a) Wide angle semi-log $\theta$-$2\theta$ XRD scan of a 20-nm thick YIG film grown on GGG (111). (b) Semi-log $\theta$-$2\theta$ scans of 10, 20, 50, and 80 nm thick YIG films near the YIG (444) peak, all of which exhibit clear Laue oscillations corresponding to the film thickness. The vertical short lines mark the positions of the YIG (444) peak. The scans are offset from each other for clarity. The insets are the rocking curves of the four YIG films taken for the first satellite peak to the left of the man peak at the $2\theta$ angle marked by the up arrows. The shoulder in the rocking curve of the 80-nm film is likely due to twinning in the film.

**Figure 2.** Room-temperature FMR derivative spectrum $dI_{FMR}/dH$ vs. $H$ of a 20-nm YIG film (YIG-1) at $\theta_H = 90°$ (field in-plane) gives a linewidth of 7.4 Oe. Top-right inset: schematic of FMR experimental geometry. Bottom-left inset: angular dependence of $H_{res}$ for the YIG film and the fit (solid green curve) agrees well with the experimental data, from which $4\pi M_{eff} = 1794$ Oe and $g = 2.0$ were obtained.

**Figure 3.** (a) Schematic diagram of the ISHE voltage measurement setup. (b) $V_{ISHE}$ vs. $H$ spectra at $\theta_H = 90°$ and $P_{rf} = 200$ mW for two Pt(5nm)/YIG(20nm) (Pt/YIG-1 and Pt/YIG-2) and a W(5nm)/YIG(20nm) (W/YIG-2) bilayers give an ISHE voltage of 1.74 mV, 2.10 mV and -5.26 mV, respectively. (c) rf power dependence of $V_{ISHE}$ with a least-squares fit for the three samples. (d) $V_{ISHE}$ vs. $H$ spectra at different $\theta_H$ for Pt/YIG-1 and W/YIG-2. The curves are offset for clarity. The non-zero ISHE voltage at $\theta_H = 0°$ and the difference in $H_{res}$ between Pt/YIG-1 and W/YIG-2 at the same $\theta_H$ are due to slight misalignment of the sample with respect to **H**. (e) Angular dependence of the normalized $V_{ISHE}$ for Pt/YIG-1 and W/YIG-2, where the red and blue curves show $\sin\theta_H$ and $-\sin\theta_H$, respectively.

**Figure 4.** FMR derivative absorption spectra of YIG thin films at $P_{rf} = 0.2$ mW before ($\Delta H_0$, blue) and after ($\Delta H_1$, red) the deposition of (a) 5-nm Pt on YIG-1 ($V_{ISHE} = 1.74$ mV), (b) 5-nm Pt on YIG-2 ($V_{ISHE} = 2.10$ mV), and (c) 5-nm W on YIG-2 ($V_{ISHE} = 5.26$ mV), which show linewidth increase from $\Delta H_0 = 7.4$ Oe to $\Delta H_1 = 27.3$ Oe, from 11.7 Oe to 36.0 Oe, and from 11.7 Oe to 24.0 Oe, respectively.

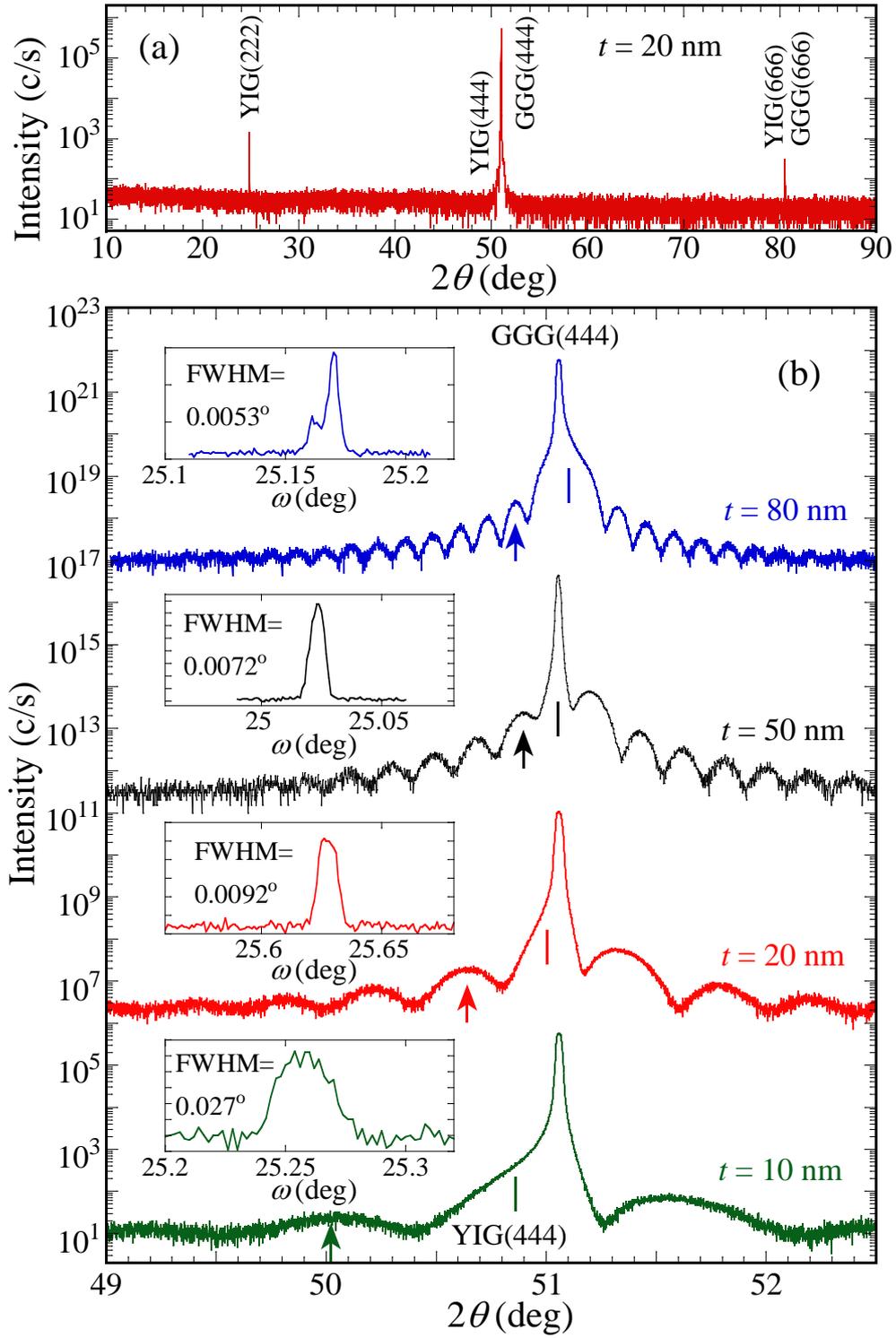

**Figure 1.**

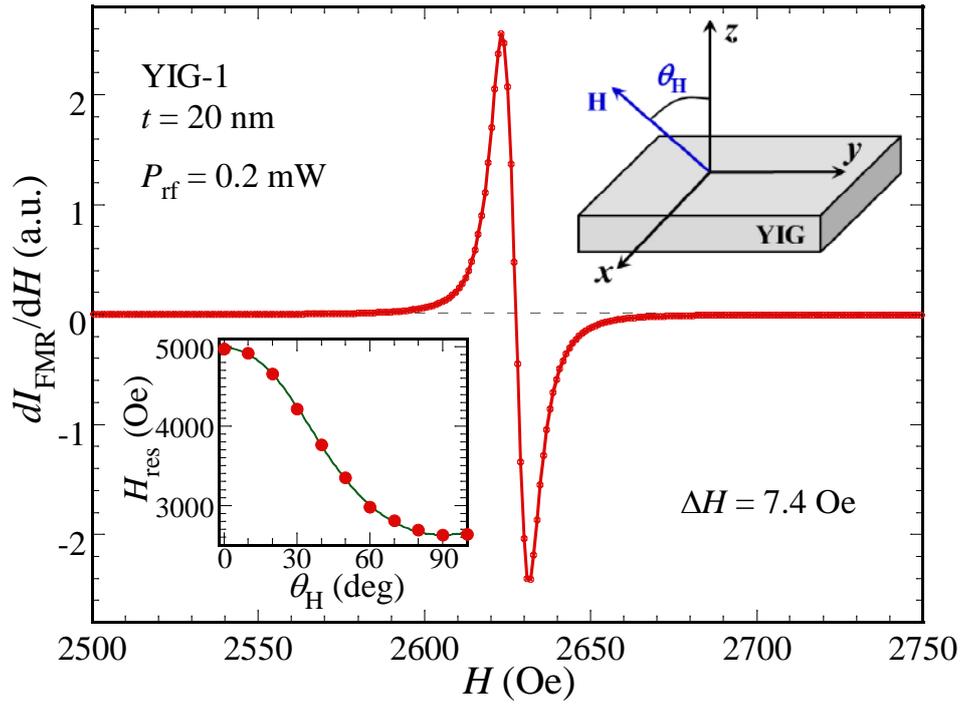

**Figure 2**.

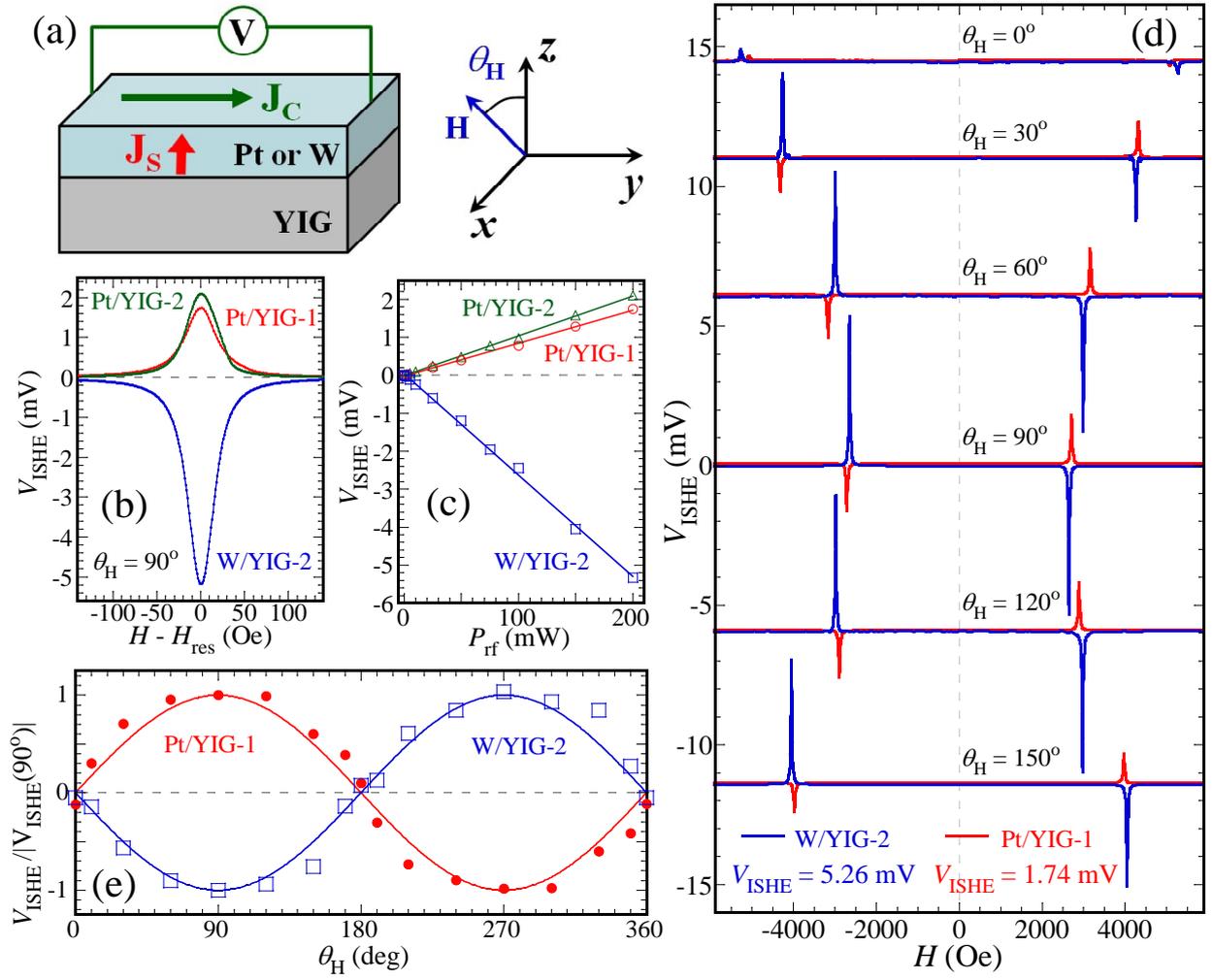

**Figure 3.**

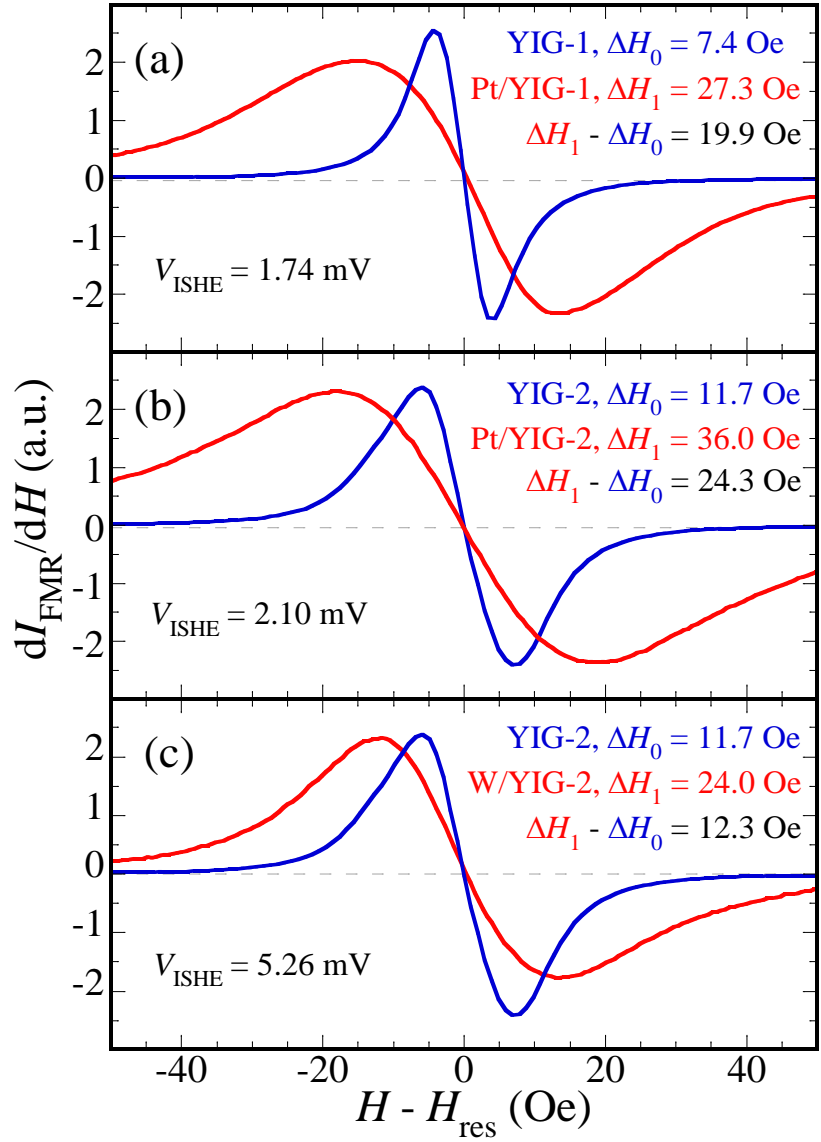

**Figure 4**.

**Supplementary Information for**

**Large Spin Pumping from Epitaxial $Y_3Fe_5O_{12}$ Thin Films to Pt and W Layers**

H. L. Wang, C. H. Du, Y. Pu, R. Adur, P. C. Hammel, and F. Y. Yang

1. **Growth of $Y_3Fe_5O_{12}$ films**

Single-crystalline $Y_3Fe_5O_{12}$ (YIG) epitaxial thin films were grown on (111)-oriented $Gd_3Ga_5O_{12}$ (GGG) substrates in an off-axis ultrahigh vacuum (UHV) sputtering system with a base pressure below $5 \times 10^{-9}$ Torr. Horizontal sputtering sources and 90° off-axis geometry were used for film deposition [S1-S3]. The optimal growth conditions include: a total Ar/$O_2$ pressure of 11.5 mTorr with an $O_2$ concentration of 0.15%, a substrate temperature 750°C, and a radio-frequency sputtering power of 50 W. The deposition rate for YIG is 0.33 nm/min and the film thickness ranges from 10 to 200 nm. For FMR and spin pumping measurements, the film thickness is typically 20 nm.

2. **Deposition of Pt and W films**

Pt and W films of 5-nm thick were deposited in the same off-axis UHV sputtering system for YIG film growth. DC magnetron sputtering was used with a deposition rate of about 1.7 nm/minute for Pt and W.

3. **Magnetization characterization by FMR**

Saturation magnetization and $g$ factor of the YIG films were determined from FMR

resonance field as a function of $\theta_H$. Resonant condition can be derived by minimizing the total free energy $F$. For a material with tetragonal symmetry [S4], $F$ can be expressed by:

$$F = -\boldsymbol{H} \cdot \boldsymbol{M} + \frac{1}{2}M\left\{4\pi M_{eff}\cos^2\theta - \frac{1}{2}H_{4\perp}\cos^4\theta - \frac{1}{8}H_{4\|}(3+\cos4\phi)\sin^4\theta - H_{2\|}\sin^2\theta\sin^2(\phi-\frac{\pi}{4})\right\}, \quad (S1)$$

where θ and ϕ are angles of magnetization (*M*) in the equilibrium position with respect to the film normal and in-plane easy axes, respectively. The first term in Eq. (S1) is the Zeeman energy and the second term is the effective demagnetizing energy ($4\pi M_{eff}$) which includes both the shape anisotropy ($4\pi M_s$) and out-of-plane uniaxial anisotropy $H_{2\perp}$, where $4\pi M_{eff} = 4\pi M_s - H_{2\perp}$. The remaining terms are out-of-plane cubic anisotropy ($H_{4\perp}$), in-plane cubic anisotropy ($H_{4\|}$), and in-plane uniaxial anisotropy ($H_{2\|}$).

The equilibrium orientation (θ, ϕ) of magnetization can be obtained by minimizing the free energy, and the FMR resonance frequency ω in equilibrium is given by [S4]:

$$\left(\frac{\omega}{\gamma}\right)^2 = \frac{1}{M^2\sin^2\theta}\left[\frac{\partial^2 F}{\partial\theta^2}\frac{\partial^2 F}{\partial\phi^2} - \left(\frac{\partial^2 F}{\partial\theta\partial\phi}\right)^2\right],$$

(S2)

where $\gamma = g\mu_B/\hbar$ is the gyromagnetic ratio. We used a numerical procedure to obtain the equilibrium angles at resonance condition at different $\theta_H$ [S5]. By fitting the data in the bottom-left inset to Fig. 2, we obtained $4\pi M_{eff}$ = 1794 Oe and *g* factor = 2.0. Given that YIG is magnetically soft, all the anisotropy terms should be small.